# Solutions of the scattering problem in a complete set of Bessel functions with a discrete index

A. D. Alhaidari[(a)] and M. E. H. Ismail[(b)]

[(a)] *Saudi Center for Theoretical Physics, P.O. Box 32741, Jeddah 21438, Saudi Arabia*

[(b)] *Department of Mathematics, University of Central Florida, Orlando, FL 32816, USA*

**Abstract:** We use the tridiagonal representation approach to solve the radial Schrödinger equation for the continuum scattering states of the Kratzer potential. We do the same for a radial power-law potential with inverse-square and inverse-cube singularities. These solutions are written as infinite convergent series of Bessel functions with a discrete index. As physical application of the latter solution, we treat electron scattering off a neutral molecule with electric dipole and electric quadrupole moments.

## 1. Introduction and TRA formulation

The tridiagonal representation approach (TRA) is an algebraic method for solving linear ordinary differential equations of the second order [1,2]. It was inspired by the J-matrix method, which was proposed many years ago by Yamani *et al* to handle the scattering problem in quantum mechanics [3-5]. Originally, the J-matrix method was considered a purely physical technique but then Ismail and his collaborators turned it into a mathematical technique [6-8]. In the TRA, the solution of the differential equation (e.g., Schrödinger equation) is written as a pointwise convergent series (finite or infinite) in terms of a complete set of square integrable functions (basis). The set is chosen such that the matrix representation of the wave operator is tridiagonal. Consequently, the differential equation becomes a three-term recursion relation for the expansion coefficients of the series. The recursion is then solved in terms of orthogonal polynomials in the space of parameters of the differential equation (e.g., the energy, angular momentum and potential parameters). All properties of the solution (e.g., energy spectrum, scattering phase shift, density of states, etc.) are obtained from the properties of these polynomials (e.g., their weight function, generating function, zeros, asymptotics, etc.). For details on how to use the TRA for solving quantum mechanical problems, one may consult [9,10] and references therein.

Our objective here is to use the TRA to solve the following time-independent Schrödinger equation (in the atomic units $\hbar = M = 1$) for a set of potentials $\{V(r)\}$ at an energy $E$ in a given special basis

$$\left[ -\frac{1}{2}\frac{d^2}{dr^2} + V(r) \right] \psi(r) = E\psi(r), \qquad (1)$$

where $r_- \leq r \leq r_+$ and $r_\pm$ are the boundaries of the physical configuration space. It should be noted that in two (or three) dimensions with cylindrical (or spherical) symmetry and $r \geq 0$, the potential function $V(r)$ stands for the effective radial potential that includes the orbital term. That is, $V(r) \mapsto \frac{L^2 - \frac{1}{4}}{2r^2} + V(r)$, where in 2D $L = m = 0, \pm 1, \pm 2, ...$ and in 3D $L = \ell + \frac{1}{2} = \frac{1}{2}, \frac{3}{2}, \frac{5}{2}, ...$. We start by expanding the solution as series of a complete set of functions $\{\phi_n(x)\}$. That is, $\psi(r) = \sum_n C_n \phi_n(x)$, where $r = f(x)$ is a coordinate transformation with $r_\pm = f(x_\pm)$ and $\{C_n\}$ are expansion coefficients that depend on the potential parameters and energy. All physical information about the system is contained in $\{C_n\}$. In general, the expansion is valid in $L^2$ but in many cases, the series is pointwise convergent. We demonstrate this numerically and show that in some special cases we recover known exact results. We write $\phi_n(x) = g(x) P_n(x)$, where $g(x)$ is a node-less (having no zeros) entire function and $\{P_n(x)\}$ is a complete set of special functions (not necessarily polynomials) that satisfy the following differential equation, differential property, and recursion relation

$$\left[ p(x) \frac{d^2}{dx^2} + q(x) \frac{d}{dx} + h(n,x) \right] P_n(x) = 0, \tag{2a}$$

$$t(x) \frac{d}{dx} P_n(x) = d_n P_n(x) + u_{n-1} P_{n-1}(x) + w_n P_{n+1}(x), \tag{2b}$$

$$s(x) P_n(x) = a_n P_n(x) + b_{n-1} P_{n-1}(x) + c_n P_{n+1}(x), \tag{2c}$$

Writing equation (1) as $\mathcal{D}\psi(r) = 0$, where $\mathcal{D}$ is the wave operator, turns it into the sum $\sum_n C_n \mathcal{D}\phi_n(x) = 0$. The action of the wave operator on $\phi_n(x)$ in terms of the new dimensionless variable $x$ reads as follows

$$\mathcal{D}\phi_n(x) = -\frac{g}{2(f')^2} \left\{ \frac{d^2}{dx^2} + \left( 2\frac{g'}{g} - \frac{f''}{f'} \right) \frac{d}{dx} + \left( \frac{g''}{g} - \frac{g'}{g} \frac{f''}{f'} \right) + 2(f')^2 [E - V(r)] \right\} P_n(x). \tag{3}$$

The prime stands for the derivative with respect to $x$. Multiplying the inside of the curly brackets by $p(x)$ makes the left factor $-\frac{g/p}{2(f')^2}$ and the differential terms inside become $p\frac{d^2}{dx^2} + p[2(g'/g) - (f''/f')]\frac{d}{dx}$. Therefore, for a given special function $P_n(x)$, we can utilize the differential equation (2a) to rewrite Eq. (3) as

$$\mathcal{D}\phi_n(x) = -\frac{g/p}{2(f')^2} \left\{ (\tilde{q} - q) \frac{d}{dx} - h(n,x) + p\left[ \frac{g''}{g} - \frac{g'}{g} \frac{f''}{f'} \right] + 2p(f')^2 [E - V(r)] \right\} P_n(x), \tag{4}$$

where $\tilde{q} = p[2(g'/g) - (f''/f')]$. We require that $\tilde{q}$ be of the same functional form as $q$. Now, the TRA dictates that $\mathcal{D}\phi_n(x)$ takes the following form [9,10]

$$\mathcal{D}\phi_n(x) = \omega(x) [\alpha_n \phi_n(x) + \beta_{n-1} \phi_{n-1}(x) + \gamma_n \phi_{n+1}(x)], \tag{5}$$

where $\{\alpha_n, \beta_n, \gamma_n\}$ are $x$-independent parameters and $\omega(x)$ is a node-less entire function. Consequently, two distinct scenarios emerge. The first and simplest is when $\tilde{q} = q$ in which case

$$2(g'/g) = (f''/f') + (q/p),\tag{6}$$

and Eq. (4) becomes

$$\mathcal{D}\phi_n(x) = -\frac{g/p}{2(f')^2}\left\{-h(n,x) + p\left[\frac{g''}{g} - \frac{g'}{g}\frac{f''}{f'}\right] + 2p(f')^2[E - V(r)]\right\}P_n(x).\tag{7}$$

The second and more elaborate scenario is if $\tilde{q} \neq q$ in which case Eq. (4) becomes

$$\mathcal{D}\phi_n(x) = -\frac{(\tilde{q}-q)g}{2(f')^2 pt}\left\{d_n P_n(x) + u_{n-1} P_{n-1}(x) + w_n P_{n+1}(x)\right.$$
$$\left. -\frac{g/p}{2(f')^2}\left[-h(n,x) + p\left(\frac{g''}{g} - \frac{g'}{g}\frac{f''}{f'}\right) + 2p(f')^2(E - V)\right]P_n(x)\right\}\tag{8}$$

where we have used the differential relation (2b). Therefore, all potential functions that are solvable using the TRA in the basis $\{\phi_n(x)\}$ must be chosen such that (7) or (8) reduces, with the help of the recursion relation (2c), to the TRA fundamental constraint (5). This means that the potential function $V(r)$ and $E$ must be chosen such that

$$\frac{1}{2p(f')^2}\left\{h(n,x) + p\left[\frac{g'}{g}\frac{f''}{f'} - \frac{g''}{g}\right] + 2p(f')^2(V - E)\right\} = \tilde{c}\,\omega(x)[s(x) + \tilde{s}],\tag{9}$$

where the constants $\tilde{c}$ and $\tilde{s}$ may depend on $n$. Additionally, for the first scenarios where $\tilde{q} = q$ we must impose (6) whereas for the second scenario where $\tilde{q} \neq q$, we must conclude that $\omega(x) \propto (q - \tilde{q})/2(f')^2 pt$.

Consequently for the two scenarios, the wave equation (1) which reads $\sum_n C_n \mathcal{D}\phi_n(x) = 0$, and after using the TRA constraint (5), becomes equivalent to the following three-term recursion relation for the expansion coefficients of the wavefunction

$$\alpha_n C_n + \beta_n C_{n+1} + \gamma_{n-1} C_{n-1} = 0,\tag{10}$$

for $n = 1, 2, 3,...$ which is solvable for all $\{C_n\}$ starting with two initial values $C_0$, $C_1$. Hence, a solution of the differential wave equation (1) is equivalent to the solution of the difference equation (10). This constitutes the foundation of the algebraic structure of the TRA as a method for solving linear ordinary differential equations.

In Section 2, we use the TRA outlined above with the special function $P_n(x)$ being chosen as a discretized Bessel function $J_{n+\nu}(z)$ and obtain the corresponding set of potentials that satisfy the TRA constraints (5). In Sections 3 and 4, we derive two associated TRA solutions of the scattering problem. In Section 5, we present an interesting physical application where we treat electron scattering off a neutral atom/molecule with electric dipole and electric quadrupole moments. Finally, in Section 6 we give the TRA solutions for two novel potential models in a basis constructed using the odd and even discrete Bessel functions defined in Appendix A.

## 2. TRA in the discrete Bessel basis

The Bessel function $J_\nu(z)$ with a general complex parameter and argument is one of the most studied special functions in the mathematical physics literature [11,12]. It appears in the solution of many problems in science and engineering. For the current problem, we choose the special function $P_n(x)$ in the TRA basis $\{\phi_n(x)\}$ as the Bessel function with a discretized index, $J_{n+\nu}(x)$. The argument $x$ and parameter $\nu$ are real such that $\nu > 0$, $x_- = 0$, and $x_+ \to +\infty$. Moreover, $n = 0,1,2,..$ and the basis parameter $\nu$ will be determined by the TRA requirements. The properties of $J_{n+\nu}(x)$ needed for the TRA are as follows [11,12]

$$\left[ x^2 \frac{d^2}{dx^2} + x \frac{d}{dx} + x^2 - (n+\nu)^2 \right] J_{n+\nu}(x) = 0, \tag{11a}$$

$$\frac{d}{dx} J_{n+\nu}(x) = \frac{1}{2} \left[ J_{n-1+\nu}(x) - J_{n+1+\nu}(x) \right], \tag{11b}$$

$$\frac{1}{x} J_{n+\nu}(x) = \frac{1/2}{n+\nu} \left[ J_{n+1+\nu}(x) + J_{n-1+\nu}(x) \right]. \tag{11c}$$

Additional relevant properties are shown in Appendix A. Moreover, to justify our expansion of the solution in these discrete Bessel functions, we needed to show that they satisfy an orthogonality on the positive real line. This is also shown in Appendix A as Eq. (A10).

The problem of polynomial expansion is a central theme in the theory of special functions. One of the oldest expansion problem is that of the plane wave $e^{ixy}$ (the kernel of the Fourier transform), which is expanded in spherical harmonics as, [see Eq. (4.8.3) in Ref. 13]

$$\exp(ixy) = \Gamma(\nu)(y/2)^{-\nu} \sum_{n=0}^{\infty} i^n (n+\nu) C_n^\nu(x) J_{n+\nu}(y),$$

where $\{C_n^\nu(x)\}$ are the ultra-spherical polynomials (spherical harmonics). Fields and Wimp [14] extended this expansion to the hypergeometric function of argument $xy$ in hypergeometric polynomials of $y$ and the coefficients are power series in $x$. One of the most general expansion of this type is, see [15],

$$\sum_{n=0}^{\infty} a_n b_n \frac{(xy)^n}{n!} = \sum_{n=0}^{\infty} \left\{ \frac{(-x)^n}{n!(n+\gamma)_n} \left( \sum_{r=0}^{\infty} \frac{b_{n+r} x^r}{r!(2n+\gamma+1)_r} \right) \left[ \sum_{s=0}^{n} \frac{(-n)_s (n+\gamma)_s}{s!} a_s y^s \right] \right\}.$$

All of these treat expansions of functions of $xy$ in functions of $x$. The expansions studied in this work, however, expand general functions of $r$ and $E$ as a sum of products of functions of $r$ times functions of $E$, where $E$ is the spectral variable. In some sense this is a separation of variables technique.

Comparing Eq. (11) with Eq. (2), we can identify the following objects: $p(x) = x^2$, $q(x) = x$, $h(n,x) = x^2 - (n+\nu)^2$, $s(x) = x^{-1}$, and $t(x) = 1$. The requirement that $\tilde{q}$ must have the same functional form as $q$ means that $2(g'/g) - (f''/f')$ is proportional to $x^{-1}$, which implies that $(f''/f') \propto x^{-1} + R(x)$ and $2(g'/g) \propto x^{-1} + R(x)$, where $R(x)$ is an arbitrary function. If we choose $R(x) = 0$, then we obtain

–4–

$$g(x) = x^{-\mu}, \qquad f'(x) := \frac{dr}{dx} = x^{\rho}/\lambda, \qquad (12)$$

where the dimensionless parameters $\rho$ and $\mu$ are to be determined by the TRA requirement and $\lambda$ is a positive scale parameter of inverse length dimension. Using (12), the action of the wave operator on the basis $\phi_n(x) = x^{-\mu} J_{n+\nu}(x)$ given by (4) becomes

$$\mathcal{D}\phi_n(x) = -\frac{\lambda^2}{2} x^{-(2\rho+\mu+2)}$$
$$\times \left\{ -(2\mu+\rho+1)x\frac{d}{dx} - x^2 + (n+\nu)^2 + \mu(\mu+\rho+1) + \frac{2}{\lambda^2} x^{2(\rho+1)}[E - V(r)] \right\} J_{n+\nu}(x) \qquad (13)$$

## 2.1 The first scenario

For the first scenario (7), we must take $2\mu + \rho + 1 = 0$, which reduces (13) to

$$\mathcal{D}\phi_n(x) = -\frac{\lambda^2}{2} x^{3\mu} \left\{ -x^2 + (n+\nu)^2 - \mu^2 + \frac{2}{\lambda^2} x^{-4\mu}[E - V(r)] \right\} J_{n+\nu}(x). \qquad (14)$$

The TRA fundamental constraint (5) and recursion relation (11c) require that the terms inside the curly brackets are either constants or proportional to $x^{-1}$ with $\omega(x) = -(\lambda^2/2)x^{4\mu}$. For energy independent potentials, we end up with three possible cases that are compatible with the TRA fundamental constraint (5):

1. $\mu = 0$: $\quad \rho = -1$, $x(r) \propto e^{\lambda r}$, $-\infty < r < +\infty$, and $V(r) = -\frac{1}{2}\lambda^2 x^2 + Ax^{-1}$. (15a)
2. $\mu = \frac{1}{4}$: $\quad \rho = -\frac{3}{2}$, $x(r) = (2/\lambda r)^2$, $r \geq 0$, and $V(r) = -\frac{1}{2}\lambda^2 x^3 + Ax$. (15b)
3. $\mu = -\frac{1}{2}$: $\quad \rho = 0$, $x(r) = \lambda r$, $r \geq 0$, $\lambda^2 = 2E$, and $V(r) = Ax^{-2} + Bx^{-3}$. (15c)

Where $A$ and $B$ are arbitrary real parameters. Another possibility in this scenario takes place if we multiply the inside of the curly brackets in (14) by $x^{-1}$ making $\omega(x) = -(\lambda^2/2)x^{4\mu+1}$ and leading to the following three additional cases as a result of the TRA fundamental constraint (5):

4. $\mu = -\frac{1}{4}$: $\quad \rho = -\frac{1}{2}$, $x(r) = (\lambda r/2)^2$, $r \geq 0$, and $V(r) = -\frac{1}{2}\lambda^2 x + Ax^{-1}$. (16a)
5. $\mu = 0$: $\quad \rho = -1$, $x(r) \propto e^{\lambda r}$, $-\infty < r < +\infty$, and $V(r) = -\frac{1}{2}\lambda^2 x^2 + Ax$. (16b)
6. $\mu = -\frac{1}{2}$: $\quad \rho = 0$, $x(r) = \lambda r$, $r \geq 0$, $\lambda^2 = 2E$, and $V(r) = Ax^{-1} + Bx^{-2}$. (16c)

Only two cases out of the above six in (15) and (16) are physically acceptable. These are (15c) and (16c). Three out of the reaming unacceptable four cases correspond to potential functions that are infinitely repulsive causing all solutions to escape to infinity (i.e., these cases do not result in local physical solutions). On the other hand, the potential function for case (15b) suffers from quantum anomalies due to the supercritical attractive singularity at the origin, $-r^{-6}$, leading to the well-known problem of "falling to the center" [16]. Moreover, reality dictates that the TRA solutions for cases (15c) and (16c) are only for positive energies. That is, we can only obtain continuum scattering states solutions for these two cases. No bound states solutions could be obtained using the TRA in the discrete Bessel function basis for these cases.

## 2.2 The second scenario

For the second scenario (8), we take $2\mu+\rho+1\neq 0$. Using the differential relation (11b) turns (13) into

$$\mathcal{D}\phi_n(x) = -\frac{\lambda^2}{2}x^{-(2\rho+\mu+1)}\left\{\left(\mu+\frac{\rho+1}{2}\right)[J_{n+1+\nu}(x)-J_{n-1+\nu}(x)]\right.$$
$$\left.+\left(-x+\frac{1}{x}\left[(n+\nu)^2+\mu(\mu+\rho+1)\right]+\frac{2}{\lambda^2}x^{2\rho+1}[E-V(r)]\right)J_{n+\nu}(x)\right\} \quad (17)$$

Thus, $\omega(x)=-(\lambda^2/2)x^{-(2\rho+1)}$ and for energy independent potentials, we end up with three cases that are compatible with the TRA fundamental constraint (5):

1. $\rho=-\frac{1}{2}$: $\quad x(r)=(\lambda r/2)^2$, $r\geq 0$, and $V(r)=-\frac{1}{2}\lambda^2 x + Ax^{-1}$. (18a)
2. $\rho=-1$: $\quad x(r)\propto e^{\lambda r}$, $-\infty<r<+\infty$, and $V(r)=-\frac{1}{2}\lambda^2 x^2+Ax$. (18b)
3. $\rho=0$: $\quad x(r)=\lambda r$, $r\geq 0$, $\lambda^2=2E$, and $V(r)=Ax^{-1}+Bx^{-2}$. (18c)

For reasons similar to those in the first scenario above, only case (18c) is physically acceptable. However, it is evident that case (18c) is physically equivalent to (16c) except that the basis parameter $\mu$ is not constrained to be $-1/2$. Therefore, we limit our TRA solution in the following two sections to the first scenario where we consider the case (16c) in Section 3 and the case (15c) in Section 4.

## 3. The Kratzer potential

In this section, we obtain the TRA scattering solution of the discrete algebraic equation (10) for the case (16c) where the potential function is the Kratzer potential [17,18]

$$V(r)=\frac{\xi}{r}+\frac{\Lambda}{r^2}, \quad (19)$$

with $A=k\xi$, $B=k^2\Lambda$, where $k=\lambda=\sqrt{2E}$ and $\Lambda>-1/8$. The basis element, $\phi_n(x)=\sqrt{kr}\,J_{n+\nu}(kr)$, becomes energy dependent. The action of the wave operator on this basis, which is given by (14) with $\mu=-1/2$ and this potential, reduces to following

$$\mathcal{D}\phi_n(x)=-\frac{E}{\sqrt{x}}\left\{\left[(n+\nu)^2-\left(2\Lambda+\frac{1}{4}\right)\right]\frac{1}{x}-2\frac{\xi}{k}\right\}J_{n+\nu}(x). \quad (20)$$

Using the recursion relation (11c) turns this into (5) with $\omega(x)=-E/2x$ and

$$\alpha_n=-4\frac{\xi}{k}, \qquad \beta_{n-1}=\gamma_n=(n+\nu)-\frac{2\Lambda+1/4}{n+\nu}. \quad (21)$$

Consequently, the recursion relation (10) for the expansion coefficients $\{C_n(E)\}$ becomes

$$zQ_n(z)=\left[(n+\nu+1)-\frac{2\Lambda+1/4}{n+\nu+1}\right]Q_{n+1}(z)+\left[(n+\nu-1)-\frac{2\Lambda+1/4}{n+\nu-1}\right]Q_{n-1}(z), \quad (22)$$

where $z := 4\xi/k$ and we wrote $C_n(E) := C_0(E)Q_n(z)$ making $Q_0(z) = 1$. This recursion relation is solvable (albeit not in closed form) for all $\{Q_n(z)\}_{n=0}^{\infty}$ starting with the two initial values $Q_{-1}(z) := 0$ and $Q_0(z) = 1$. Finally, the scattering wavefunction becomes

$$\psi(r) = C_0(E)\sqrt{kr}\sum_{n=0}^{\infty} Q_n(4\xi/k) J_{n+\nu}(kr). \qquad (23)$$

The solution of the Coulomb problem, which is a special case is presented below as an example, suggests that the basis parameter $\nu$ must be chosen as $\nu^2 = 2\Lambda + \frac{1}{4}$. Consequently, the polynomials $Q_n(z)$ will not be orthogonal because their recursion relation (22) violates the spectral theory requirement that $\beta_n \gamma_n > 0$ for all $n$ since $\gamma_0 = 0$ [13]. Additionally, $\{Q_n(z)\}_{n=1}^{\infty}$ have a common root at $z = 0$. On the other hand, if we define the polynomials $V_n(z)$ by $Q_n(z) = \frac{z(n+\nu)}{n} V_{n-1}(z)$ for $n \geq 1$, then $\{V_n(z)\}_{n=0}^{\infty}$ will be a set of orthogonal polynomials that satisfy the following recursion relation

$$z(n+\nu+1)V_n(z) = (n+1)(n+2\nu+2)V_{n+1}(z) + (n+1)(n+2\nu)V_{n-1}(z), \qquad (24)$$

where $V_{-1}(z) = 0$ and $V_0(z) = (\nu+1)/(2\nu+1)$. In Appendix B, we derive some of the analytic properties of these orthogonal polynomials including their weight function, generating function and asymptotics ($n \to \infty$). Accordingly, the wavefunction expansion (23) could be rewritten in terms of these orthogonal polynomials as follows

$$\psi(r) = C_0(E)\sqrt{kr}\left[J_\nu(kr) + 4\frac{\xi}{k}\sum_{n=0}^{\infty}\left(1 + \frac{\nu}{n+1}\right)V_n(4\xi/k) J_{n+\nu+1}(kr)\right]. \qquad (25)$$

The energy factor $C_0(E)$ is determined by the boundary condition. For example, we may adopt the condition

$$\lim_{r\to\infty}\psi(r) = \cos(kr + \delta). \qquad (26)$$

Then, using the well-known asymptotic behavior (as $x \to \infty$) of the Bessel function $J_{n+\nu}(x)$, we obtain from (23)

$$\lim_{r\to\infty}\psi(r) = \sqrt{\frac{2}{\pi}} C_0(E)\sum_{n=0}^{\infty} Q_n(4\xi/k)\cos\left[kr - \left(n+\nu+\tfrac{1}{2}\right)\tfrac{\pi}{2}\right]. \qquad (27)$$

Equation (26) and (27) give the following two relations

$$\tan[\delta(E)] = -\mathcal{S}(E)/\mathcal{C}(E), \qquad (28a)$$

$$C_0^2(E) = \frac{\pi/2}{\mathcal{S}^2(E) + \mathcal{C}^2(E)}, \qquad (28b)$$

where $\mathcal{S}(E) := \sum_{n=0}^{\infty} \sin[(n+\nu+\tfrac{1}{2})\tfrac{\pi}{2}] Q_n(4\xi/k)$ and $\mathcal{C}(E) := \sum_{n=0}^{\infty} \cos[(n+\nu+\tfrac{1}{2})\tfrac{\pi}{2}] Q_n(4\xi/k)$.

To test the validity of our solution, we consider the special case of Coulomb scattering where $\Lambda = \frac{1}{2}\ell(\ell+1)$, $\nu = \ell + \frac{1}{2}$ and $\xi = Z$ is the electric charge. On the other hand, the Coulomb wavefunction is well-known and written as [11,19]

$$\psi(r) = \frac{2^\ell e^{-\pi\sigma/2}}{\Gamma(2\ell+2)} |\Gamma(\ell+1\pm i\sigma)| (kr)^{\ell+1} e^{\pm ikr} \, {}_1F_1(\ell+1\pm i\sigma; 2\ell+2; \mp 2ikr), \qquad (29)$$

where $\sigma = Z/k$ and ${}_1F_1(a;b;z)$ the confluent hypergeometric function. For a given value of the energy $E$, angular momentum $\ell$ and electric charge $Z$, Figure 1 shows an excellent match of the Coulomb wavefunction (29) with the TRA solution (23) where the recursion relation (22) becomes

$$4\sigma Q_n(4\sigma) = \left[(n+\ell+\tfrac{3}{2}) - \frac{(\ell+\tfrac{1}{2})^2}{n+\ell+\tfrac{3}{2}}\right] Q_{n+1}(4\sigma) + \left[(n+\ell-\tfrac{1}{2}) - \frac{(\ell+\tfrac{1}{2})^2}{n+\ell-\tfrac{1}{2}}\right] Q_{n-1}(4\sigma). \qquad (30)$$

An interesting byproduct of our findings above is a new representation of the confluent hypergeometric function as an infinite sum of the discrete Bessel functions $J_{n+\nu}(z)$ [20].

Due to the long-range interaction of the $r^{-1}$ term in the Kratzer potential (19), the asymptotic behavior of the wavefunction (23) is not exactly as shown by Eq. (26) but, in fact, it reads $\cos[kr - (\xi/k)\ln(2kr) + \delta]$. Therefore, the expressions derived for the phase shift $\delta(E)$ and energy factor $C_0(E)$ are not as given by Eq. (28). Nonetheless, the exact Coulomb wave function (29) and the representation of ${}_1F_1(a;b;z)$ obtained in [20] give

$$C_0(E) = \frac{\sqrt{\pi/2}}{\Gamma(\ell+1)} e^{-\pi\sigma/2} |\Gamma(\ell+1\pm i\sigma)| = \frac{\sqrt{\pi/2}}{\Gamma\left(\tfrac{1}{2}+\sqrt{2\Lambda+\tfrac{1}{4}}\right)} e^{-\pi\xi/2k} \left|\Gamma\left(\tfrac{1}{2} + \sqrt{2\Lambda+\tfrac{1}{4}} \pm i\tfrac{\xi}{k}\right)\right|. \qquad (31)$$

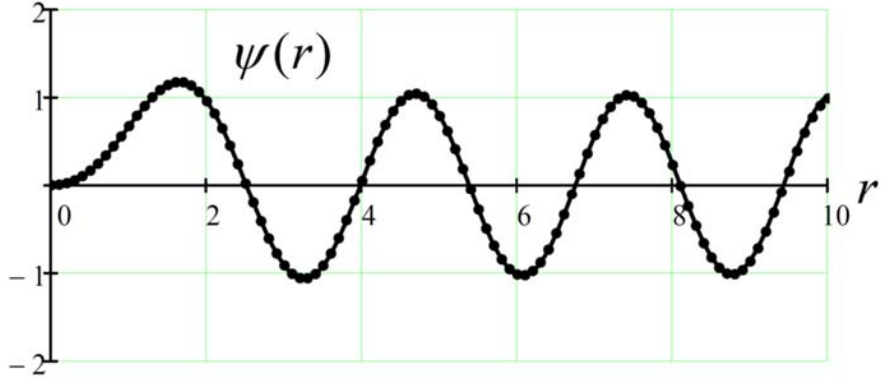

**Fig. 1**: The Coulomb scattering wavefunction (dotted curve) overlaid by the TRA solution (solid curve). We took $Z = 2$, $\ell = 1$ and $E = 3.0$ (in atomic units). The horizontal axis $r$ is measured in units of the Bohr radius.

## 4. Inverse power-law potential

In this section, we obtain the TRA scattering solution (10) for the case (15c) where the potential function is the following power-law potential with inverse square and inverse cube singularities at the origin

$$V(r) = \frac{\Lambda}{r^2} + \frac{\zeta}{r^3},  \quad (32)$$

where $A = k^2\Lambda$ and $B = k^3\zeta$. The basis element is $\phi_n(x) = \sqrt{kr}\, J_{n+\nu}(kr)$ and the action of the wave operator on this basis, which is given by (14) with $\mu = -1/2$ and this potential, reduces to following

$$\mathcal{D}\phi_n(x) = -Ex^{-3/2}\left[(n+\nu)^2 - \left(2\Lambda + \tfrac{1}{4}\right) - 2k\zeta x^{-1}\right] J_{n+\nu}(x). \quad (33)$$

Using the recursion relation (11c) turns this into (5) with $\omega(x) = -E/x^2$ and

$$\alpha_n = (n+\nu)^2 - \left(2\Lambda + \tfrac{1}{4}\right), \qquad \beta_{n-1} = \gamma_n = \frac{-k\zeta}{n+\nu}. \quad (34)$$

Consequently, the recursion relation (10) for the expansion coefficients $\{C_n(E)\}$ becomes

$$z\left[(n+\nu)^2 - \left(2\Lambda + \tfrac{1}{4}\right)\right] Q_n(z) = \frac{Q_{n+1}(z)}{n+\nu+1} + \frac{Q_{n-1}(z)}{n+\nu-1}. \quad (35)$$

where $z = 1/k\zeta$ and we wrote $C_n(E) := C_0(E) Q_n(z)$ making $Q_0(z) = 1$. This recursion relation is solvable for all $\{Q_n(z)\}_{n=0}^{\infty}$ starting with the two initial values $Q_{-1}(z) := 0$ and $Q_0(z) = 1$. The scattering wavefunction becomes

$$\psi(r) = C_0(E)\sqrt{kr} \sum_{n=0}^{\infty} Q_n(1/k\zeta) J_{n+\nu}(kr). \quad (36)$$

In analogy with the solution of the Kratzer problem above, we fix the arbitrary basis parameter $\nu$ as $\nu^2 = 2\Lambda + \tfrac{1}{4}$ giving $Q_1(z) = 0$, $\nu Q_2(z) = -(\nu+2)$ and $\nu Q_3(z) = -4(\nu+1)(\nu+2)(\nu+3)z$. As a result, $Q_n(z)$ will be a polynomial in $z$ of degree $n-2$, which suggests that we define another polynomial as $W_n(z) = \frac{-\nu Q_{n+2}(z)}{n+\nu+2}$. Thus, $W_0(z) = 1$, $W_1(z) = 4(\nu+1)(\nu+2)z$ and (35) becomes the following recursion relation for the orthogonal polynomials $\{W_n(z)\}$

$$(n+2)(n+2+\nu)(n+2+2\nu)\, z W_n(z) = W_{n+1}(z) + W_{n-1}(z), \quad (37)$$

The monic form of these polynomials $\{\tilde{W}_n(z)\}$ satisfy the following recursion relation

$$z\tilde{W}_n(z) = \tilde{W}_{n+1}(z) + \frac{\tilde{W}_{n-1}(z)}{(n+1)_2 (n+\nu+1)_2 (n+2\nu+1)_2}.$$

Therefore, the corresponding tridiagonal matrix operator is a bounded compact operator and the orthogonality measure is purely discrete and bounded. Let $(-A, A)$ be the smallest interval containing the support of the measure. Using chain sequences, [13] it is easy to see that

$$A \leq \frac{1}{3(\nu+1)_3 (2\nu+1)_3}.$$

Finally, the wavefunction series (36) could be written in terms of these orthogonal polynomials as follows

$$\psi(r) = C_0(E)\sqrt{kr}\left[ J_\nu(kr) - \sum_{n=0}^{\infty}\left(1 + \tfrac{n+2}{\nu}\right) W_n(1/k\zeta) J_{n+\nu+2}(kr) \right]. \tag{38}$$

In the following section, we apply these findings to electron scattering off a neutral molecule that has a permanent electric dipole and quadrupole moments.

## 5. Electron scattering off a molecule with electric dipole and quadrupole moments

As physical application of the solution obtained in Section 4, we study electron scattering off a neutral atom/molecule with electric dipole and quadrupole moments.

In spherical coordinates, the electrostatic potential function $V(\vec{r})$ associated with such a neutral molecule becomes a multipole expansion, which is written up to and including the linear electric quadrupole as follows (see, for example, Ref. [21])

$$V(\vec{r}) = -d\frac{\cos\theta}{r^2} + q\frac{\tfrac{1}{2}(3\cos^2\theta - 1)}{r^3}, \tag{39}$$

where $d$ is the electric dipole moment along the positive $z$-axis, and $q$ is the linear electric quadrupole moment of the molecule. We took the Bohr radius, $a_0 = 4\pi\varepsilon_0 \hbar^2 / Me^2 = 4\pi\varepsilon_0 / e^2$, as the unit of length with $M$ and $-e$ being the mass and charge of the electron. Now, the quadrupole term in (39) destroys separability of the wave equation making its solution a highly non-trivial task. Therefore, we consider an effective electric quadrupole interaction where the angular factor $\tfrac{1}{2}(3\cos^2\theta - 1)$ is replaced by a dimensionless angular parameter $\eta$ such that $-\tfrac{1}{2} \leq \eta \leq +1$ since $0 \leq \cos^2\theta \leq 1$. This results in an effective quadrupole potential $p/r^3$, where the effective electric quadrupole moment is $p = \eta q$. Consequently, the radial part of the wave equation becomes

$$\left[ -\frac{1}{2}\frac{d^2}{dr^2} + \frac{\chi(\chi+1)}{2r^2} + \frac{p}{r^3} - E \right]\psi(r) = 0, \tag{40}$$

where $\chi$ is a quantum number that depends on the electric dipole moment $d$ and the azimuthal angular momentum quantum number $m = 0, \pm 1, \pm 2, \dots$ [22]. If $d = 0$ then $\chi$ becomes the orbital angular momentum quantum number $\ell = 0, 1, 2, \dots$. On the other hand, for a non-zero dipole moment $d$, Eq. (3.6) and Eq. (3.8) in Ref. [22] give the value of $\left(\chi+\tfrac{1}{2}\right)^2$ as one of the eigenvalues of an infinite symmetric tridiagonal matrix whose elements are

$$T_{i,j} = \left(i+m+\tfrac{1}{2}\right)^2 \delta_{i,j} - d\sqrt{\tfrac{i(i+2m)}{(i+m)^2-1/4}}\,\delta_{i,j+1} - d\sqrt{\tfrac{(i+1)(i+2m+1)}{(i+m+1)^2-1/4}}\,\delta_{i,j-1}, \tag{41}$$

where $i, j = 0, 1, 2, \ldots$. Reality dictates that $d$ must be less than a certain critical value $d_{max}$, otherwise $|m|$ must be chosen greater than a minimum integer $m_{min}$.

Using Eq. (36), we obtain the electron scattering wavefunction as follows

$$\psi(r) = C_0(E)\sqrt{kr}\sum_{n=0}^{\infty}\left(1+\tfrac{n}{\nu}\right)\tilde{Q}_n(z)J_{n+\nu}(kr), \tag{42}$$

where $\tilde{Q}_n(z) = \dfrac{\nu}{n+\nu}Q_n(z)$ and

$$n(n+\nu)(n+2\nu)z\tilde{Q}_n(z) = \tilde{Q}_{n+1}(z) + \tilde{Q}_{n-1}(z), \tag{43}$$

where $z = 1/kp$, $\nu = \chi + \tfrac{1}{2}$, $\tilde{Q}_0(z) = 1$, $\tilde{Q}_1(z) = 0$, $\tilde{Q}_2(z) = -1$ and $\tilde{Q}_3(z) = -4(\nu+1)(\nu+2)z$. Now, since the effective potential in this problem, which reads $V(r) = \tfrac{1}{2}\chi(\chi+1)r^{-2} + pr^{-3}$, does not contain a long-range component, then we can assume that the asymptotic behavior of the corresponding wavefunction is given by Eq. (26). Consequently, the phase shift and wavefunction energy factor $C_0(E)$ must be given by Eq. (28), where

$$S(E) := \sum_{n=0}^{\infty} \sin[(n+\nu+\tfrac{1}{2})\tfrac{\pi}{2}]\left(1+\tfrac{n}{\nu}\right)\tilde{Q}_n(1/kp), \tag{44a}$$

$$C(E) := \sum_{n=0}^{\infty} \cos[(n+\nu+\tfrac{1}{2})\tfrac{\pi}{2}]\left(1+\tfrac{n}{\nu}\right)\tilde{Q}_n(1/kp), \tag{44b}$$

For a given electric dipole moment $d$ and azimuthal quantum number $m$, we use one of the eigenvalues of the matrix whose elements are given by (41) to obtain an angular quantum number $\chi$. Figure 2 is a plot of the wavefunction (42) at a given energy for the physical parameters $d$, $\chi$, $q$ and with a given choice of the angular parameter $\eta$.

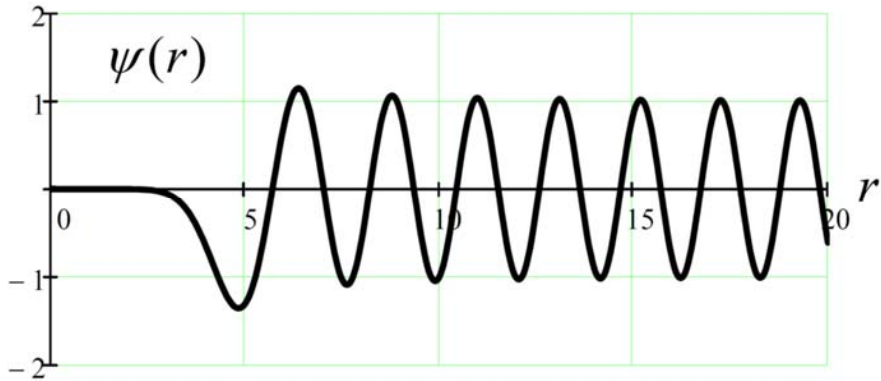

**Fig. 2**: Wavefunction of an electron scattering off a neutral molecule whose physical parameters and energy (in atomic units) are $\{d,q,m,E\} = \{2,3,1,5\}$. The angular parameter was taken as $\eta = \frac{1}{2}$ and the horizontal $r$ axis is measured in units of the Bohr radius.

## 6. Two additional novel potential models

Using a basis written in terms of the odd and even discrete Bessel functions $\mathcal{J}_n^\nu(x)$ and $\mathcal{K}_n^\nu(x)$ defined in Appendix A, we obtain TRA solutions for two interesting and unique potential models. Following the same procedure as in Section 2, we start by evaluating the action of the wave operator on the basis then apply the TRA fundamental constraint (5). The results are stated below without giving details.

### 6.1 The one-dimensional exponential potential

In the 1D square-integrable odd basis $\phi_n(x) = \mathcal{J}_n^\nu(e^{\lambda r})$ [23] where $-\infty < r < +\infty$, $\mu = 0$ and $\rho = -1$, we obtain bound state solutions for the following exponential potential

$$V(r) = V_0 e^{2\lambda r}, \tag{45}$$

where $V_0 = -\lambda^2/2$. These correspond to diagonal rather than tridiagonal representations of the wave operator where the solutions are written as $\psi_n(r) = C_n \phi_n(x) = \sqrt{2(2n+\nu+1)}\mathcal{J}_n^\nu(e^{\lambda r})$. The corresponding energy spectrum is

$$E_n = -\frac{\lambda^2}{2}(2n+\nu+1)^2. \tag{46}$$

In the even basis $\phi_n(x) = \mathcal{K}_n^\nu(e^{\lambda r})$, similar result is obtained where $\psi_n(r) = \sqrt{2(2n+\nu)}\mathcal{K}_n^\nu(e^{\lambda r})$ and the corresponding energy spectrum is $E_n = -\frac{1}{2}\lambda^2(2n+\nu)^2$. However, the potential (45) is highly singular (exponentially repulsive at infinity) causing quantum anomalies and requiring potential regularization or self-adjoint extension of the Hamiltonian [24-30]. This anomaly is evident by noting that the energy spectrum is bottomless (extending all the way to minus infinity) and that each bound state has an infinite number of nodes with rapidly increasing spatial oscillations as $r \to +\infty$. These anomalous features come about despite the fact that the bound states are square integrable (normalizable). Similar anomalous features occur for the inverse squared potential with supercritical coupling [30].

### 6.2 The inverse quartic radial potential

In the energy-dependent odd basis $\phi_n(r) = \sqrt{kr}\mathcal{J}_n^\nu(kr)$ where $\mu = -1/2$ and $\rho = 0$, we obtain the scattering solution for the following radial power-law singular potential

$$V(r) = \frac{\Lambda}{r^2} + \frac{\zeta}{r^4}, \tag{47}$$

–12–

where $\zeta > 0$. These solutions are written as $\psi(r) = C_0(E) \sum_{n=0}^{\infty} Q_n \phi_n(r)$, where $\{Q_n\}$ satisfy the following three-term recursion relation

$$\left(2\Lambda + \tfrac{1}{4}\right) Q_n = \left[(2n+\nu+1)^2 - \frac{\zeta k^2}{(2n+\nu+1)^2 - 1}\right] Q_n$$
$$- \frac{\zeta k^2}{2}\left[\frac{Q_{n+1}}{(2n+\nu+2)(2n+\nu+3)} + \frac{Q_{n-1}}{(2n+\nu)(2n+\nu-1)}\right] \quad (48)$$

This recursion gives all $\{Q_n\}$ starting with the two initial values $Q_{-1} := 0$ and $Q_0 = 1$. In the even basis $\phi_n(r) = \sqrt{kr} \mathcal{K}_n^\nu(kr)$, similar results are obtained.

## Acknowledgements

We are grateful to H. Bahlouli for fruitful discussions. The support provided by SCTP is highly appreciated.

## Data Availability Statement:

Data sharing is not applicable to this article as no new data were created or analyzed in this study.

## Appendix A. Discrete Bessel functions: odd, even and orthogonality

The following integral is a special case of that in Section 3.8.5 on page 99 of the book [12]:

$$\int_0^\infty x^{-\mu} J_{n+\nu}(x) J_{m+\nu}(x) dx =$$
$$2^{-\mu} \Gamma(\mu) \Gamma\left(\tfrac{1-\mu}{2} + \nu + \tfrac{n+m}{2}\right) \Big/ \Gamma\left(\tfrac{1+\mu}{2} + \nu + \tfrac{n+m}{2}\right) \Gamma\left(\tfrac{1+\mu}{2} + \tfrac{n-m}{2}\right) \Gamma\left(\tfrac{1+\mu}{2} - \tfrac{n-m}{2}\right) \quad (A1)$$

where $n + m + 2\nu + 1 > \mu > 0$. If $n = m$, we obtain

$$\int_0^\infty x^{-\mu} [J_{n+\nu}(x)]^2 dx = \frac{\Gamma(\mu)/2^\mu}{\left[\Gamma\left(\tfrac{1+\mu}{2}\right)\right]^2}\left[\Gamma\left(\tfrac{1-\mu}{2} + \nu + n\right) \Big/ \Gamma\left(\tfrac{1+\mu}{2} + \nu + n\right)\right]$$
$$= \frac{2^\mu}{4\pi} \frac{[\Gamma(\mu/2)]^2}{\Gamma(\mu)} \left[\Gamma\left(\tfrac{1-\mu}{2} + \nu + n\right) \Big/ \Gamma\left(\tfrac{1+\mu}{2} + \nu + n\right)\right] \quad (A2)$$

In the integral formula (A1), the expression $\left[\Gamma\left(\tfrac{1+\mu}{2} + \tfrac{n-m}{2}\right)\Gamma\left(\tfrac{1+\mu}{2} - \tfrac{n-m}{2}\right)\right]^{-1}$ with $n \neq m$ vanishes if $|n - m| = \mu + 1, \mu + 3, \mu + 5, \ldots$. Therefore, for non-negative integers $n$ and $m$, $\mu$ must be a positive integer for that to happen (i.e., $\mu := k = 1, 2, 3, \ldots$). Making a table of the integral (A1) for a fixed $v$ and a given $k$, we can show that (A1) becomes an orthogonality (i.e., proportional to $\delta_{n,m}$) in the following two scenarios:

(1) $k$ even: $n$ must be odd and $m$ even or vice versa but with $|n-m| \geq k+1$.

(2) $k$ odd: both $n$ and $m$ are even or both are odd but with $|n-m| \geq k+1$.

However, the orthogonality will be of size $k$ (i.e., the orthogonality matrix is band-diagonal with a band size of $k$), which dictates that $k=1$. Moreover, this choice of $k$ will keep the same symmetry for the set of integers $n$ and $m$ (i.e., both even or both odd). Hence, we obtain the following orthogonality

$$\int_0^\infty x^{-1} J_{2n+v}(x) J_{2m+v}(x) dx = \frac{1/2}{2n+v} \delta_{n,m}, \tag{A3a}$$

$$\int_0^\infty x^{-1} J_{2n+1+v}(x) J_{2m+1+v}(x) dx = \frac{1/2}{2n+v+1} \delta_{n,m}. \tag{A3b}$$

where $n,m = 0,1,2,3,\ldots$. For $m \neq n$, one can show that the orthogonality (A3) follows directly from the Sturm-Liouville problem of the Bessel function. On the other hand,

$$\int_0^\infty x^{-1} J_{2n+v}(x) J_{2m+1+v}(x) dx = \frac{(-1)^{n+m}}{2\pi \left(m+n+v+\frac{1}{2}\right)\left(m-n+\frac{1}{2}\right)}, \tag{A3c}$$

Moreover, taking the limit $\mu \to 0$ in (A1) gives

$$\int_0^\infty J_{2n+v}(x) J_{2m+1+v}(x) dx = \sigma_{n,m} \frac{1}{2}, \tag{A3d}$$

where $\sigma_{n,m}$ is a $\pm$ sign that depends on $n$ and $m$ as follows

$$\sigma_{n,m} = \begin{cases} (-1)^{n+m} & , n \leq m \\ -(-1)^{n+m} & , n > m \end{cases}$$

Hence, we define $\mathcal{K}_n^v(x) := J_{2n+v}(x)$ and $\mathcal{J}_n^v(x) := J_{2n+v+1}(x) = \mathcal{K}_n^{v+1}(x) = \mathcal{K}_{n+\frac{1}{2}}^v(x)$ to obtain from the recursion relation (11c) the following three-term recursion relations:

$$\frac{2}{x^2} \mathcal{K}_n^v(x) = \frac{1}{(2n+v)^2-1} \mathcal{K}_n^v(x)$$
$$+ \frac{1/2}{(2n+v)(2n+v-1)} \mathcal{K}_{n-1}^v(x) + \frac{1/2}{(2n+v)(2n+v+1)} \mathcal{K}_{n+1}^v(x) \tag{A4a}$$

$$\frac{2}{x^2} \mathcal{J}_n^v(x) = \frac{1}{(2n+v+1)^2-1} \mathcal{J}_n^v(x)$$
$$+ \frac{1/2}{(2n+v)(2n+v+1)} \mathcal{J}_{n-1}^v(x) + \frac{1/2}{(2n+v+1)(2n+v+2)} \mathcal{J}_{n+1}^v(x) \tag{A4b}$$

where $\mathcal{K}_0^v(x) = J_v(x)$, $\mathcal{K}_1^v(x) = J_{v+2}(x)$, $\mathcal{J}_0^v(x) = J_{v+1}(x)$, and $\mathcal{J}_1^v(x) = J_{v+3}(x)$. We also obtain from (11b) and (11c) the following two differential properties

–14–

$$\frac{2}{x}\frac{d}{dx}\mathcal{K}_n^\nu(x) = \frac{1}{(2n+\nu)^2-1}\mathcal{K}_n^\nu(x) + \frac{1/2}{2n+\nu-1}\mathcal{K}_{n-1}^\nu(x) - \frac{1/2}{2n+\nu+1}\mathcal{K}_{n+1}^\nu(x) \tag{A5a}$$

$$\frac{2}{x}\frac{d}{dx}\mathcal{J}_n^\nu(x) = \frac{1}{(2n+\nu+1)^2-1}\mathcal{J}_n^\nu(x) + \frac{1/2}{2n+\nu}\mathcal{J}_{n-1}^\nu(x) - \frac{1/2}{2n+\nu+2}\mathcal{J}_{n+1}^\nu(x) \tag{A5b}$$

Additionally, we obtain from (A3) the following orthogonality and cross integrals

$$\int_0^\infty x^{-1}\mathcal{K}_n^\nu(x)\mathcal{K}_m^\nu(x)dx = \frac{1/2}{2n+\nu}\delta_{n,m}, \tag{A6a}$$

$$\int_0^\infty x^{-1}\mathcal{J}_n^\nu(x)\mathcal{J}_m^\nu(x)dx = \frac{1/2}{2n+\nu+1}\delta_{n,m}. \tag{A6b}$$

$$\int_0^\infty x^{-1}\mathcal{K}_n^\nu(x)\mathcal{J}_m^\nu(x)dx = \frac{(-1)^{n+m}}{2\pi\left(m-n+\frac{1}{2}\right)\left(n+m+\nu+\frac{1}{2}\right)}, \tag{A6c}$$

$$\int_0^\infty \mathcal{K}_n^\nu(x)\mathcal{J}_m^\nu(x)dx = \sigma_{n,m}\frac{1}{2}. \tag{A6d}$$

The discrete Bessel functions has the following property (see Section 6.5 in [13])

$$J_{n+\nu}(z) = h_{n,\nu}(1/z)J_\nu(z) - h_{n-1,\nu+1}(1/z)J_{\nu-1}(z), \tag{A7}$$

where $h_{n,\nu}(z)$ are the Lommel polynomials which are generated from the initial values $h_{0,\nu}(z)=1$ and $h_{1,\nu}(z)=2\nu z$ by the recursion

$$2z(n+\nu)h_{n,\nu}(z) = h_{n+1,\nu}(z) - h_{n-1,\nu}(z), \tag{A8}$$

Let $0 < j_{\nu,1} < j_{\nu,2} < ... < j_{\nu,n} < ...$ be the positive zeros of $J_\nu(z)$, then Lommel polynomials have the following discrete orthogonality [13]

$$\sum_{k=1}^\infty \frac{1}{j_{\nu,k}^2}\left[h_{n,\nu+1}(1/j_{\nu,k})h_{m,\nu+1}(1/j_{\nu,k}) + h_{n,\nu+1}(-1/j_{\nu,k})h_{m,\nu+1}(-1/j_{\nu,k})\right] = \frac{\delta_{n,m}}{2(n+\nu+1)}. \tag{A9}$$

It is clear from (A7) that

$$J_{n+\nu+1}(j_{\nu,k}) = h_{n,\nu}(1/j_{\nu,k})J_{\nu+1}(j_{\nu,k})$$

Therefore, the orthogonality relation (A9) becomes

$$\left[1+(-1)^{n+m}\right]\sum_{k=1}^\infty \frac{J_{n,\nu+1}(j_{\nu,k})J_{m,\nu+1}(j_{\nu,k})}{j_{\nu,k}^2 J_{\nu+1}^2(j_{\nu,k})} = \frac{\delta_{n,m}}{2(n+\nu+1)}, \tag{A10}$$

where in the last step, we used the fact that

$$J_{n,\nu+1}(-z)J_{m,\nu+1}(-z) = (-1)^{n+m}J_{n,\nu+1}(z)J_{m,\nu+1}(z).$$

# Appendix B. Orthogonal polynomials associated with scattering in the Kratzer problem

The recurrence relation (24) is a special case of the most general recurrence relation for which this technique works that reads

$$x(n+a)P_n(x) = (n+1)(n+b)P_{n+1}(x) + (n+\alpha)(n+\beta)P_{n-1}(x), \tag{B1}$$

with $P_0(x) = 1$ and $P_1(x) = ax/b$. The canonical form of this recursion for the monomial polynomials $y_n(x)$ is

$$x\, y_n(x) = y_{n+1}(x) + \frac{n(n+b-1)(n+\alpha)(n+\beta)}{(n+a)(n+a-1)} y_{n-1}(x). \tag{B2}$$

Therefore, the general case has four free parameters satisfying the positivity condition

$$\frac{n(n+b-1)(n+\alpha)(n+\beta)}{(n+a)(n+a-1)} > 0, \quad \text{for } n > 0.$$

Let $G(x,t) = \sum_{n=0}^{\infty} P_n(x) t^n$. Multiplying both sides of (B1) by $t^n$ and summing, turns it into the following differential equation for $G(x,t)$

$$t(1+t^2)G_{tt} + \left[b - xt + (\alpha+\beta+3)t^2\right]G_t + \left[(\alpha+1)(\beta+1)t - ax\right]G = 0. \tag{B3}$$

This is a Heun equation with four regular singular points at $t = 0, i, -i, \infty$ [31]. To put this equation in the standard form we let $u = it$ and the differential equation (B3) becomes

$$G_{uu} + \left[\frac{b}{u} + \frac{\alpha+\beta+3-b+ix}{2(u-1)} + \frac{\alpha+\beta+3-b-ix}{2(u+1)}\right]G_u - \frac{iax + u(\alpha+1)(\beta+1)}{u(1-u^2)}G = 0. \tag{B4}$$

The Frobenius exponents at $t = 0, i, -i$ are

$$(0, 1-b), \quad \left(0, \tfrac{1}{2}[1+b-\alpha-\beta-ix]\right), \quad \left(0, \tfrac{1}{2}[1+b-\alpha-\beta+ix]\right),$$

respectively. Therefore, the equation has only one solution which is analytic at $t = 0$, provided that $b$ is not an integer. The generating function $G$ will then be this solution. The same conclusion holds for the $t = \pm i$. We shall apply Darboux' method to the generating function (see Chapter 8 of [32]). Let $U(x,t)$ be the singular solution at $t = i$. We then find the connection constant $A(x)$ such that $G(x,t) - A(x)U(x,t)$ be analytic in $|t| < 1$. Similarly, $B(x)$ is the connection constant at $t = -i$. Therefore the dominant terms in a comparison function is

$$A(x)(1+it)^{(1+b-\alpha-\beta-ix)/2} + B(x)(1-it)^{(1+b-\alpha-\beta+ix)/2}.$$

The coefficient of $t^n$ in the above function is

$$i^n \frac{\left(\tfrac{1}{2}[\alpha+\beta-b-1+ix]\right)_n}{n!} A(x) + (-i)^n \frac{\left(\tfrac{1}{2}[\alpha+\beta-b-1-ix]\right)_n}{n!} B(x). \tag{B5}$$

Since the polynomials are real valued, then $B(x) = \overline{A(x)}$. Using

$$\frac{(a)_n}{n!} = \frac{\Gamma(n+a)}{\Gamma(a)\Gamma(n+1)}, \qquad \frac{\Gamma(n+a)}{\Gamma(n+b)} \approx n^{a-b}.$$

where the symbol $\approx$ stands for the asymptotic ($n \to \infty$) limit. Therefore, we conclude that

$$P_n(x) \approx \frac{i^n n^{(\alpha+\beta-b-3+ix)/2}}{\Gamma\left(\frac{1}{2}[\alpha+\beta-b-1+ix]\right)} A(x) + \text{complex conjugate}. \tag{B6}$$

Therefore the scattering amplitude is

$$2|A(x)| n^{(\alpha+\beta-b-3)/2} \Big/ \left|\Gamma\left(\tfrac{1}{2}[\alpha+\beta-b-1+ix]\right)\right|.$$

This leads to the weight function

$$\omega(x) = \frac{\left|\Gamma\left(\tfrac{1}{2}[\alpha+\beta-b-1+ix]\right)\right|^2}{4|A(x)|^2}. \tag{B7}$$

For the special case of the Kratzer problem in Section 3, the four parameters are:

$$b = 2a = 2(\nu+1), \qquad \alpha = 1, \qquad \beta = 2\nu.$$

Therefore, the corresponding weight function becomes

$$\omega(x) = \frac{|\Gamma(ix/2)|^2}{(x^2+4)|A(x)|^2}, \tag{B8}$$

where the $\nu$ dependence is implicit in $A(x)$.